\documentclass[final]{svjour3}
\usepackage[dvipdfmx]{graphicx}
\usepackage{rotating}
\usepackage{amssymb}
\usepackage{mathptmx}
\usepackage[numbers]{natbib}
\makeatletter
\journalname{Journal of Low Temperature Physics}

\bibpunct{}{}{,}{s}{}{,}

\begin{document}

\newcommand{\hdblarrow}{H\makebox[0.9ex][l]{$\downdownarrows$}-}

\title{Mitigating the effects of charged particle strikes on TES arrays for exotic atom X-ray experiments}

\author{H.~Tatsuno$^{1}$
\and D.A.~Bennett$^{2}$ \and W.B.~Doriese$^{2}$ \and M.S.~Durkin$^{2}$ \and J.W.~Fowler$^{2}$ 
\and J.D.~Gard$^{2}$ \and T.~Hashimoto$^{3}$ \and R.~Hayakawa$^{1}$ \and T.~Hayashi$^{4}$ \and G.C.~Hilton$^{2}$ \and Y.~Ichinohe$^{5}$ \and H.~Noda$^{6}$ \and G.C.~O'Neil$^{2}$ \and S.~Okada$^{7}$ \and C.D.~Reintsema$^{2}$ \and D.R.~Schmidt$^{2}$ \and D.S.~Swetz$^{2}$ \and J.N.~Ullom$^{2}$ \and S.~Yamada$^{1}$ \and (the J-PARC E62 Collaboration)
}

\institute{
\begin{enumerate}
\item Department of Physics, Tokyo Metropolitan University, Tokyo 192-0397, Japan\\
\email{hideyuki.tatsuno@gmail.com}
\item National Institute of Standards and Technology, Boulder, CO 80305, USA
\item Japan Atomic Energy Agency (JAEA), Tokai 319-1184, Japan
\item Department of High Energy Astrophysics, Institute of Space and Astronautical Science (ISAS),\\
    Japan Aerospace Exploration Agency (JAXA), Kanagawa 229-8510, Japan
\item Department of Physics, Rikkyo University, Tokyo 171-8501, Japan
\item Department of Earth and Space Science, Osaka University, Osaka 560-0043, Japan
\item Atomic, Molecular and Optical Physics Laboratory, RIKEN, Wako 351-0198, Japan
\end{enumerate}
}

\maketitle

\begin{abstract}
Exotic atom experiments place transition-edge-sensor (TES) microcalorimeter arrays in a high-energy charged particle rich environment. When a high-energy charged particle passes through the silicon substrate of a TES array, a large amount of energy is deposited and small pulses are generated across multiple pixels in the TES array due to thermal crosstalk. We have developed analysis techniques to assess and reduce the effects of charged particle events on exotic atom X-ray measurements. Using this technique, the high-energy and low-energy components of the X-ray peaks due to pileup are eliminated, improving the energy resolution from 6.6 eV to 5.7 eV at 6.9 keV.


\end{abstract}

\section{Introduction}
Exotic atom X-ray spectroscopy is a unique research technique to investigate the strong interaction between negatively charged particles and a nucleus at the low-energy limit. The interaction of a kaon ($K^-$) and a nucleus is strongly attractive which shifts and broadens the electromagnetic energy levels of a kaonic atom. Therefore, precision X-ray spectroscopy of kaonic atoms has been performed to measure the interaction [1]. We have performed the J-PARC E62 experiment to measure the kaonic helium $3d\to2p$ X-rays at the J-PARC K1.8BR beamline (Ibaraki, Japan) [2]. The stopped $K^-$ in the liquid helium target forms a kaonic helium atom, then the $K^-$ cascades down to lower energy levels while emitting characteristic X-rays. The X-rays of  $3d\to2p$ transition ($\sim 6$ keV) are detected with the 240-pixel TES array, which was previously used for pionic atom X-ray spectroscopy [3]. Our goal is to determine the $2p$-state strong-interaction shift with a precision of 0.2 eV. To achieve this precision, the excellent energy resolution of TESs is essential.

The application of TESs in the charged-particle-beam environment requires non-trivial analysis techniques. One of the important analyses is to investigate the charged particle impacts on the TES array. The energy deposits of charged particles on the array, especially on its silicon substrate, can cause small thermal crosstalk pulses in all TESs. The pileup of the thermal crosstalk and normal X-ray pulses degrades the energy resolution due to poor pulse-height estimation via optimal filtering, moreover additional low-energy and high-energy tail components are needed to fit the X-ray peaks [4]. In this paper, we study the effects of charged particle events on the TES array and present analysis methods to improve the energy resolution and the energy calibration for kaonic atom X-ray spectroscopy.

\section{Charged particle event identification}
We use a group trigger to aid in the identification of thermal crosstalk from charged particle events, a feature recently added to the data acquisition software of the J-PARC E62 experiment [5,6]. In the experiment, the characteristic X-rays of pure metals (Cr, Co, and Cu) were continuously generated with an X-ray tube for the energy calibration, and the charged-particle beam at the J-PARC K1.8BR beam line was extracted for $\sim 2$ s in the $5.2$ s cycle. Thus, we have measured the calibration X-rays in both beam-on and beam-off environments.

\begin{figure}[htbp]
\centering
\includegraphics[width=0.95\linewidth, keepaspectratio]{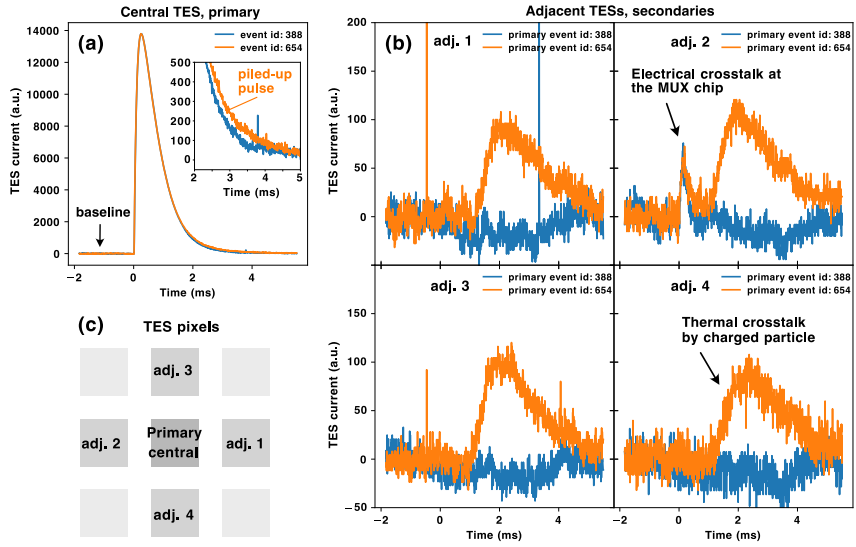}
\caption{(a) Co $K_{\alpha}$ X-ray pulses detected by the central TES (primary record). The inset shows a close-up of the region with thermal crosstalk. (b) Secondary records from the four adjacent TESs taken using the group trigger. The thermal crosstalk is seen in all adjacent TESs at 1 ms after the primary X-ray pulse. The electrical crosstalk seen in one of the adjacent TESs (adj.~2) occurs because the central TES and the adjacent TES are physically nearest neighbors on the multiplexer chip. (c) The schematic drawing of the TES-pixel positions. (Color figure online.)}
\label{fig1}
\end{figure}

In a self trigger system, a TES current pulse is sampled in every 7.2 \textrm{$\mu$}s. In the group trigger, a TES pulse triggers the recording of that TES (primary records) and the four adjacent TESs (secondary records). Fig.~\ref{fig1} (a) shows the primary record of a 6.9 keV Co $K_{\alpha}$ X-ray event that is affected by thermal crosstalk from a charged particle event (id: 654) and one that is not (id: 388), with the difference highlighted in the inset. Fig.~\ref{fig1} (b) shows the secondary records of the adjacent TESs, and Fig.~\ref{fig1} (c) shows the schematic drawing of pixel positions on the array. The thermal crosstalk can be seen in all four adjacent TESs. In this case, the charged particle hit the array 1 ms after the primary X-ray pulse, and the number of charged particle strikes is $\sim 40$ counts per second per array. Although the pulse height of the thermal crosstalk is less than 1\% of that of Co $K_{\alpha}$ X-ray, the pulse-shape difference directly affects the pulse-height determination. We discuss in the next section how to reduce the influence of this thermal crosstalk. The electrical crosstalk seen in one of the adjacent TESs (adj.~2) occurs because the central TES and the adjacent TES are physically nearest neighbors on the multiplexer (MUX) chip. 

\section{Analysis techniques to reduce the effects of thermal crosstalk}
In this section, we discuss the effects of thermal crosstalk on the X-ray energy measurement and techniques to reduce these effects in the analysis process. Due to the use of an optimal filter to determine the pulse height [7,8], the shift of pulse height of an X-ray event is strongly dependent on the arrival time of crosstalk during the record length. If thermal crosstalk arrives during the pre-trigger region (before the X-ray arrival) or in the post peak region (long after the X-ray arrival), the filtered pulse height will be decreased. If the thermal crosstalk arrives during the X-ray pulse peak region (almost the same arrival timing as X-ray), the filtered pulse height will be increased. Depending on the arrival region, the effects of thermal crosstalk from charged particle strikes can be reduced by using event cuts and by reducing record lengths.

\begin{figure}[htbp]
\centering
\includegraphics[width=0.95\linewidth, keepaspectratio]{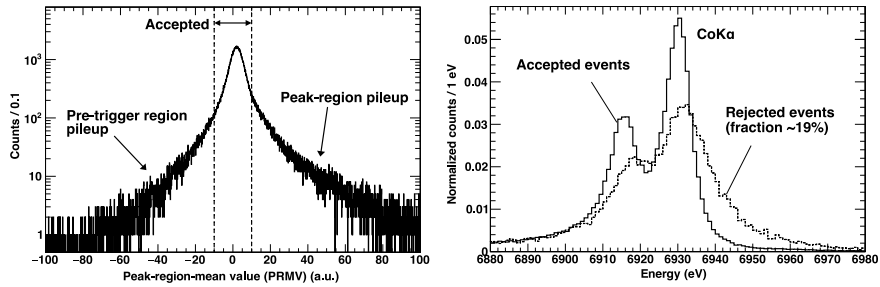}
\caption{\textit{Left}, the typical distribution of the \textit{PRMV} of the adjacent TESs. Abnormally negative and positive \textit{PRMV} indicate the pre-trigger-region pileup and the pulse-peak-region pileup, respectively. \textit{Right}, the normalized beam-on energy spectra of Co $K_{\alpha}$ X-rays summed over 221 TES pixels accepted (\textit{solid}) and rejected (\textit{dashed}) by the condition of $|\textit{PRMV}| \le 10$. The fraction of the rejected events is $\sim 19$\%.}
\label{fig2}
\end{figure}

Thermal crosstalk in the pulse peak region is identified using secondary event records from the four adjacent TES pixels. Here, we define the pulse peak region as 256 samples ($\sim 1.8$ ms) from the primary trigger and evaluate the peak-region-mean value (\textit{PRMV}) of secondary records for each adjacent TES except the nearest neighbor channel of the MUX chip to avoid the electrical crosstalk. The unit of \textit{PRMV} is the same as the sampled TES current. If there is no thermal crosstalk, the \textit{PRMV} is close to zero because it is near the baseline (the baseline is defined as the mean value of the sampling points in the pre-trigger region). A negative \textit{PRMV} indicates that the baseline of the adjacent TES is affected by the pre-trigger-region pileup, and a positive \textit{PRMV} indicates the pileup in the pulse-peak region, respectively. Fig.~\ref{fig2} shows the distribution of the \textit{PRMV} of the adjacent TESs and the Co $K_{\alpha}$ X-ray spectra accepted/rejected by the condition of $|\textit{PRMV}| \le 10$. The fraction of the rejected events is $\sim 19$\%. The high-energy component of the X-ray peak is excluded by this cut. No additional high-energy component is needed to fit the X-ray peaks.

\begin{figure}[htbp]
\centering
\includegraphics[width=0.9\linewidth, keepaspectratio]{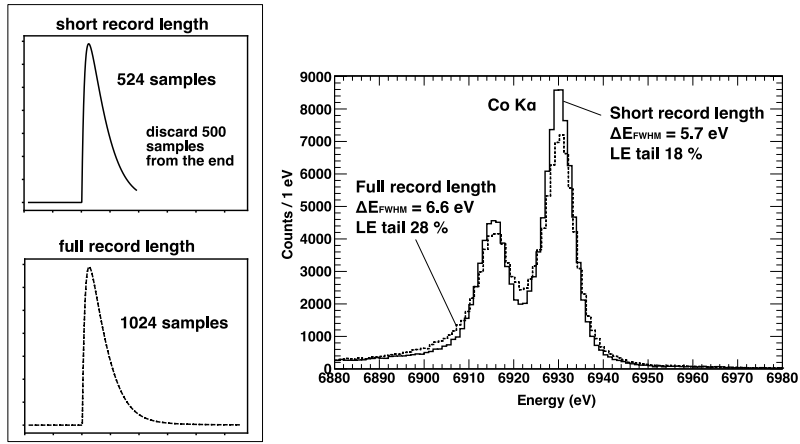}
\caption{The effects of shortening record length from 1024 samples to 524 samples. \textit{Left}, the pulse record lengths used for the analysis. \textit{Right}, the beam-on energy spectra of Co $K_{\alpha}$ X-rays summed over 221 TES pixels with the analysis of short record length (\textit{solid}) and the full record length (\textit{dashed}). The cut by \textit{PRMV} is already applied. The use of the shorter record length improves the FWHM energy resolution from 6.6 eV to 5.7 eV and reduced the low-energy tail (LE tail) fraction from 28\% to 18\%.}
\label{fig3}
\end{figure}

Thermal crosstalk in the post peak region can be eliminated by reducing record lengths. The results of this technique are shown in Fig.~\ref{fig3}, a comparison of a full record-length (1024-sample) analysis to a reduced record-length (524-sample) analysis where 500 samples are discarded at the end. By removing the last 500 samples so that thermal crosstalk in this region won't affect the pulse height determination, the full-width-at-half-maximum (FWHM) energy resolution at the Co $K_{\alpha}$ X-ray energy is improved from 6.6 eV to 5.7 eV and the fraction of the low-energy tail (LE-tail) component is also improved from 28\% to 18\%. The 18\% fraction of the low-energy tail is same with the data taken in the beam-off condition. The origin of the low-energy tail is considered as heat traps by the grain structure of the thermally evaporated bismuth absorber [9].

\section{Record Length Dependence}
As shown in Fig.~\ref{fig3}, the shorter record-length analysis is useful for the piled-up pulses in the post peak region. It improves not only the energy resolution but also the fraction of low-energy tail component without losing any events. In order to optimize the record length, we check here the record length dependence of the energy resolution. There are two parameters to shorten the record length, one is the pre-trigger region which changes from the start, the other is the post trigger region which changes from the end. We call the cut for the pre-trigger region as \textit{pre cut} and for the post trigger region as \textit{post cut}. The unit of cut is one point of waveform sampling.

\begin{figure}[htbp]
\centering
\includegraphics[width=1.00\linewidth, pagebox=cropbox, keepaspectratio]{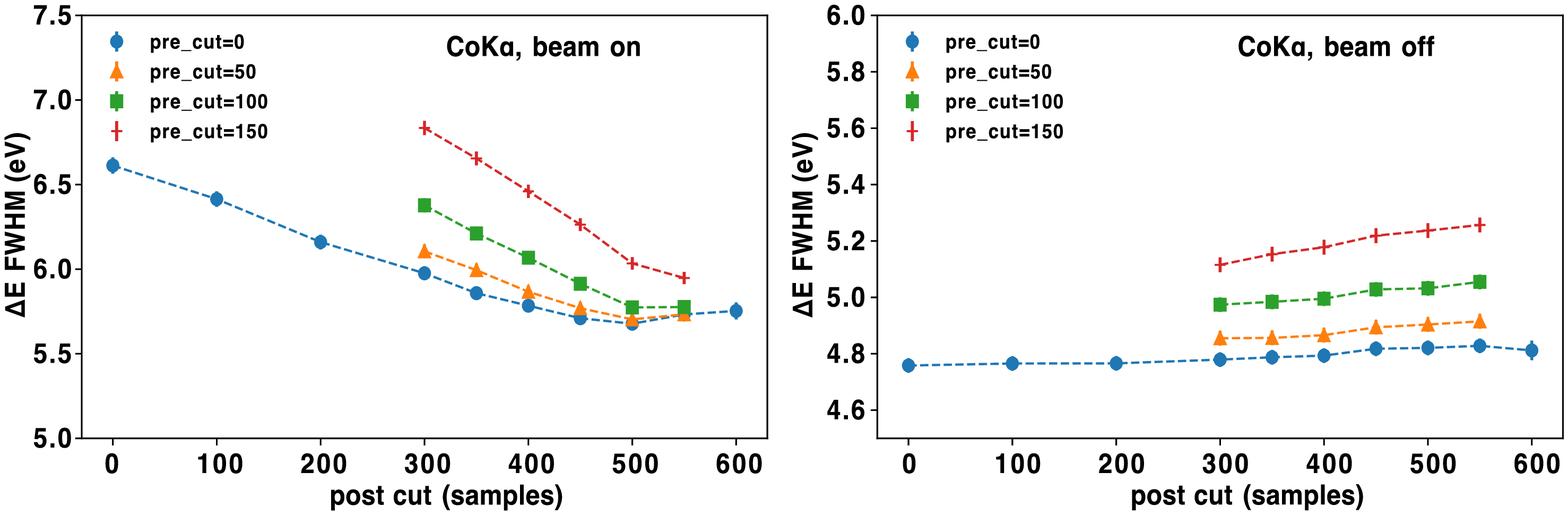}
\caption{The record length dependence of the energy resolution at the Co $K_{\alpha}$ X-ray energy as a function of \textit{post cut} samples. \textit{Left}, the beam-on data summed over 221 TES pixels, the markers \textit{circle}, \textit{triangle}, \textit{square}, and \textit{cross} show the different \textit{pre cut} of 0, 50, 100, and 150 samples, respectively. \textit{Right}, the beam-off data. The best resolution is obtained with the \textit{pre cut} $= 0$, \textit{post cut} $= 500$ point for the beam-on data. For the beam-off data, the energy resolution is degraded as the record length gets shorter. (Color figure online.)}
\label{fig4}
\end{figure}

Figure \ref{fig4} shows the beam-on and beam-off energy resolution at the Co $K_{\alpha}$ X-ray energy as a function of \textit{post cut} length with different \textit{pre cut} values. The energy resolution of beam-on data is worse than that of beam-off data because of the poor pulse-height estimation of the piled-up X-ray pulses and the base temperature fluctuation of TESs by the charged particle hits. For the beam-on data, the energy resolution improves as \textit{post cut} gets larger, and it is minimized around 500 samples. On the other hand, as \textit{pre cut} gets larger, the energy resolution gets worse. This indicates the short pre-trigger record length also affects the pulse-height estimation via optimal filtering. For the beam-off data, as both parameters get larger, the energy resolution is degraded. This looks normal for the non-piled-up pulses, because the longer record length is better to reduce the noise contribution [10]. 

\section{Analysis techniques for energy calibration of beam-on X-ray spectra}
\begin{figure}[htbp]
\centering
\includegraphics[width=1.00\linewidth]{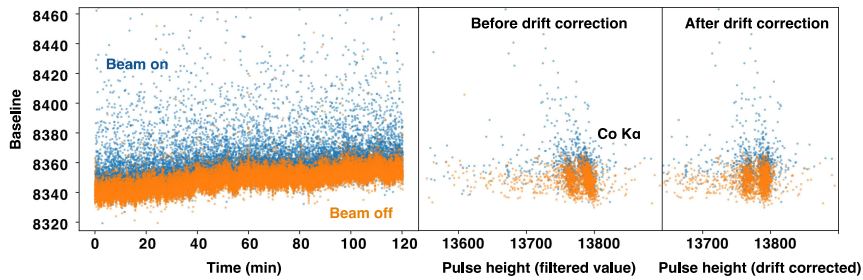}
\caption{Gain drift of the TES in 2 hours and the drift correction for the beam-on and beam-off data. The beam-on data (\textit{blue}) show the slightly higher baseline and lower pulse height than the beam-off data (\textit{orange}). (Color figure online.)}
\label{fig5}
\end{figure}

As described in the previous sections, the group-trigger analysis can eliminate the high-energy component of the X-ray pulse height spectra, and the record-length analysis can mitigate the effects on the low-energy component. The cleaned spectra are essential to minimize the systematic errors of energy calibration. The energy calibration curve is determined by the X-ray energies and the peak positions obtained by fitting the pulse height spectra. Therefore, it is important to exclude the unexpected components to obtain accurate peak positions. The data analysis procedure can be summarized as follows:
\begin{enumerate}
\item Calculate the optimal filter with the beam-off data to exclude the pileup of thermal crosstalk. If beam-off data are not available, select clean pulses with severe cut conditions.
\item Identify charged-particle hit events and optimize the record length in order to mitigate the effects of pileups with the thermal crosstalk.
\item Calibrate the energy of X-ray pulses independently for the beam-on and beam-off data. The baseline is slightly changing when the beam is on in addition to the gain drift, because the heat from the energy deposits of charged particles heats the TES array. Fig.~\ref{fig5} shows the gain drifts of the central TES for the beam-on and beam-off data, and the drift correction for the Co $K_{\alpha}$ X-ray events. The beam-on data show the slightly higher baseline than the beam-off data. In our case, the difference of the X-ray peak position between the beam-on and the beam-off data with shared drift correction is about 0.2 eV.
\item Fit each X-ray peak with the low-energy tail function (e.g., [4]) if the thermally evaporated bismuth absorbers are used. The fitted peak position without the low-energy tail is off by about 0.3 eV in our case. If the additional high-energy and low-energy components are needed, the additional systematic uncertainties should be considered.
\end{enumerate}

\section{Summary}
We have developed analysis techniques to mitigate the effects of charged particle hits on the TES array. The high-energy and low-energy components of the X-ray spectra due to pileups with the thermal crosstalk are excluded by using the pulse-peak-region analysis with the group trigger and the shorter record-length analysis to discard the piled-up record timing, respectively. These methods improve the energy resolution from 6.6 eV (FWHM) to 5.7 eV at 6.9 keV. These techniques can clean the X-ray pulse-height spectra and can minimize the systematic uncertainties of energy calibration.

\begin{acknowledgements}
This work was partly supported by the Grants-in-Aid for Scientific Research (KAKENHI) from MEXT and JSPS (Nos. 16H02190, 15H05438, 18H03714, and 18H05458). We thank the members in the NIST Quantum Sensors Project. We appreciate the significant contributions by J-PARC, RIKEN, and those who kindly have backed up the J-PARC E62 experiment.
\end{acknowledgements}

\pagebreak

\end{document}